\documentclass[12pt]{article}
\usepackage{amssymb}
\usepackage{epsfig}
\usepackage[english]{babel}
\usepackage{float}
\usepackage{slashed}
\newcommand{\be}{\begin{equation}}
\newcommand{\ee}{\end{equation}}
\newcommand{\bea}{\begin{eqnarray}}
\newcommand{\eea}{\end{eqnarray}}

\def\p{\partial}
\def\pslash{\p\raise.3ex \hbox{\kern-.5em /}}
\def\delslash{\nabla\raise.3ex \hbox{\kern-.7em /}}

\begin{document}

\vskip 5cm

\begin{center}
\Large{ \textbf{Exact Solutions for Non-Hermitian Dirac-Pauli
Equation  in an intensive  magnetic field }}
\end{center}
\vskip 0.5cm\begin{center} \Large{V.N.Rodionov}
\end{center}
\vskip 0.5cm
\begin{center}
{Plekhanov Russian University, Moscow, Russia,  \em E-mail
vnrodionov@mtu-net.ru}
\end{center}

\begin{center}

\abstract{The modified Dirac-Pauli equations, which are introduced
by means of ${\gamma_5}$-mass factorization of the ordinary
Klein-Gordon operator, are considered. We also take into account
the interaction of fermions with the intensive homogenous magnetic
field focusing attention to their (g-2) gyromagnetic factor. The
basis of this approach is developing of methods for study of the
structure of regions of unbroken $\cal PT$ symmetry of
Non-Hermitian Hamiltonians which be no studied earlier. For that,
without the use of perturbation theory in the external field the
exact energy spectra are deduced with regard to spin effects of
fermions. We also investigate the unique possible of experimental
observability the non-Hermitian restrictions in the spectrum of
mass consistent with the conjecture Markov about Maximal Mass.
This, in principal will may allow to find out the existence of an
upper limit value in spectrum masses of elementary particles and
confirm or deny the significance of the Planck mass.}

\end{center}


\section{Introduction}

 Now it is well-known fact, that the reality of the
spectrum in models with a non-Hermitian Hamiltonian is a
consequence of $\cal PT$-invariance of the theory, i.e. a
combination of spatial and temporary parity of the total
Hamiltonian: $[H,{\cal PT}]\psi =0$. When the $\cal PT$ symmetry
is unbroken, the spectrum of the quantum theory is real. This
surprising results explain the growing interest in this problem
which was initiated by Bender and Boettcher's observation
Ref.\cite{ben}. For the past a few years has been studied a lot of
new non-Hermitian $\cal PT$-invariant systems Ref.\cite{ft12} -
\cite{neznamov2}.

The non-Hermitian ${\cal PT}$-symmetric $\gamma_5$-extension of
the Dirac equation is first studied in Ref.\cite{ft12} and further
illustrated in Ref.\cite{ft13} and Ref.\cite{ManGup}-\cite{RodKr}.
The purpose of this paper is the continuation of the studying
examples of pseudo-Hermitian relativistic  Hamiltonians,
investigations of which was started  by us earlier
Ref.\cite{Rod1}-\cite{RodKr}.

 Here we are producing our
investigation of non-Hermitian systems with $\gamma_5$-mass term
extension taking into account the external magnetic field. We also
are studying the spectral and polarization properties of such
systems and with this aim we start to consider solutions of the
modified Dirac equation for free fermions. After that we take into
account interaction with the intensive magnetic fields of charged
and neutral particles having anomalous magnetic moments (AMM). The
novelty of developed by us approach is associated with predictions
of new phenomena caused by a number of additional terms of the
non-Hermitian Hamiltonians, which radically changes the picture of
interactions.  But it is not only refers to the processes with
large energies, but  may be observed in the region of low energies
when one takes into account the interaction AMM of  fermions with
 intensive magnetic fields.

This paper has the following structure. In section II the
non-Hermitian approach to the construction of plane wave solutions
is formulated for the case  free massive particles. In the third
section we study the basic characteristics of modified Dirac
models with $\gamma_5$-massive contributions in the external
magnetic field. Then, in the fourth section, we consider the
modified  Dirac-Pauli model  in the magnetic field with
non-Hermitian extension.  This section also contains the
discussion of  the effects of the possible observability of the
parameters $m_1, m_2$ taking into account the interaction of
charged fermions together with regard to their AMM with the
magnetic field. The fifth section contains Summary and
Conclusions.

\section{Non-Hermitian extensions of plane waves}

Let us now consider the solutions of modified Dirac equations for
free massive particles using the  ${\gamma_5}$-factorization of
the ordinary Klein-Gordon operator. In this case similar to the
Dirac procedure one can represent the Klein-Gordon operator in the
form of a product of two commuting matrix operators:

\be\label{D2} \Big({\partial_\mu}^2 +m^2\Big)=
\Big(i\partial_\mu\gamma^{\mu}-m_1-\gamma_5 m_2 \Big)
\Big(-i\partial_\mu\gamma^{\mu}-m_1+\gamma_5 m_2 \Big), \ee where
designations $\hbar=c=1 $ are used and the physical mass of
particles $m$ is expressed through the parameters $m_1$ and $m_2$
\be \label{012} m^2={m_1}^2- {m_2}^2. \ee

For so the function would obeyed to the equations of Klein-Gordon
\be\label{KG} \Big({\partial_\mu}^2
+m^2\Big)\widetilde{\psi}(x,t)=0 \ee one can demand that it also
satisfies  one of equations of the first order \be\label{ModDir}
\Big(i\partial_\mu\gamma^{\mu}-m_1-\gamma_5 m_2
\Big)\widetilde{\psi}(x,t)
=0\,\,\Big(-i\partial_\mu\gamma^{\mu}-m_1+\gamma_5 m_2 \Big)
\widetilde{\psi}(x,t)=0 \ee

Equations (\ref{ModDir}) of course, are less common than (\ref{KG}
), and although every solution of one of the equations
(\ref{ModDir}) satisfies (\ref{KG}), reverse approval has not
designated. It is also obvious that the Hamiltonians, associated
with the equations (\ref{ModDir}), no are Hermitian, because in it
the $\gamma_5$-dependent mass components appear ($H\neq H^{+}$):

  \be\label{H} H =\overrightarrow{\alpha} \textbf{p}+ \beta(m_1
+\gamma_5 m_2)\ee  and \be\label{H+} H^+ =\overrightarrow{\alpha
}\textbf{p}+ \beta(m_1 -\gamma_5 m_2).\ee Here  matrices
$\alpha_i=\gamma_0\cdot\gamma_i$, $\beta=\gamma_0$,
$\gamma_5=-i\gamma_0\gamma_1\gamma_2\gamma_3$.

  It is easy to see from (\ref{012}) that the  mass $m$,
appearing in the equation (\ref{KG}) is real, when the inequality
\be \label{e210} {m_1}^2\geq {m_2}^2.\ee is accomplished.

A.Mustafazadeh identified the necessary and sufficient
requirements of reality  of eigenvalues for pseudo-Hermitian and
$\cal PT$-symmetric Hamiltonians and formalized the use these
Hamilton operates  in his papers Ref.\cite{alir}, Ref.\cite{ali}
and Ref.\cite{spec}-\cite{most5}. According to the recommendations
of this works we can define Hermitian operator $\eta$, which
transform non-Hermitian Hamiltonian
 by means of invertible
transformation to the Hermitian-conjugated one. It is easy to see
that with Hermitian operator \be \label{et1} \eta= e^{\gamma_5
\vartheta}, \ee where $\vartheta= \textmd{arctanh} (m_2/m_1)$  we
can obtain
 \be\label{D3}  \eta H \eta^{-1}= H^{+},\ee

In addition, multiplying the Hamilton operator $H$ from left to $
e^{{\vartheta\gamma_5 }/2}$ and taking into account that matrices
$\gamma_5$ commute with matrices $\alpha_i$ and anti-commute with
$\beta$, we can obtain
 \be \label{H0}  e^{{\gamma_5 \vartheta}/2}
H = H_0  e^{{\gamma_5 \vartheta}/2},\ee where $H_0 =\alpha p
+\beta m $ is a ordinary Hermitian Hamiltonian of a free particle.

The mathematical sense of the action of the operator (\ref{et1})
it turns out, if we notice that according to  the properties of
$\gamma_5$ matrices, all the even  degree of $\gamma_5$ are equal
to 1, and all odd degree are equal to $\gamma_5$. Given that
$\cosh(x)$ decomposes on even and $\sinh(x)$ odd degrees of $x$,
the expressions (\ref{D3})-(\ref{H0}) can be obtained by
representing non-unitary exponential operator $\eta$ in the form
\be\label{eta} \eta=e^{\gamma_5 \vartheta}=\cosh\vartheta+\gamma_5
\sinh\vartheta,\ee  where \be\label{chsh} \cosh\vartheta =
m_1/m;\,\,\,\, \sinh\vartheta = m_2/m. \ee

 The region of the unbroken
$\cal PT$-symmetry of (\ref{H}) can  be found  in the form
(\ref{e210}). However, it is not apparent that the area with
undisturbed $\cal PT$-symmetry defined by such a way does not
include  the regions, corresponding to the  some unusual
particles, description of which radically distinguish from
traditional one.



If we used the standard representation of $\gamma$-matrixes the
non-Hermitian Hamiltonian $H$ can be whiten in the following
matrix form
$$H=\left(%
\begin{array}{cccc}
  m_1 & 0 & p_3-m_2 & p_1-ip_2 \\
  0 & m_1 & p_1+ip_2 & -m_2-p_3 \\
  m_2+p_3 & p_1-ip_2 & -m_1 & 0 \\
  p_1+ip_2 & m_2-p_3 & 0 & -m_1 \\
\end{array}%
\right),$$ where $p_i$ are components of momentum.

 It is clear that $$H\widetilde{\psi}=E\widetilde{\psi}.$$
 The condition $\det{(H-E)}=(-E^2 + {m_1}^2-{m_2}^2 + {p_{\bot}}^2 +{p_3}^2)^2 =0$
results in the eigenvalues of $E$ which are represented in the
form: \be\label{E} \emph{E}=\pm\sqrt{{m_1}^2-{m_2}^2 +
{p_{\bot}}^2 +{p_3}^2},\ee where $p_{\bot}=\sqrt{{p_1}^2+{p_2}^2}$
and ${m_1}^2-{m_2}^2 =m^2 $, that coincide with the eigenvalues of
energy  of Hermitian operator $H_0$.

Let us now consider the state of a free particle with certain
values of the momentum and energy, which is described by a plane
wave and can be written as \be\label{psi} \widetilde{\psi}=
\frac{1}{\sqrt{2E}}\widetilde{u}e^{-ipx}. \ee It is easy to see
that the wave amplitude $\widetilde{u} $ is determined by
bispinor, normalization of which now needs an additional
explanation.

Really using (\ref{ModDir}) and taking into account properties of
matrices $\gamma_o, \vec{\gamma}, \gamma_5$, we can  write also
complex-conjugate equation \be\label{conj}
\left(-p_0\tilde{\gamma_0}
-\textbf{p}\widetilde{\vec{\gamma}}-m_1-\gamma_5
m_2\right)\widetilde{\psi^{*}}=0, \ee where $\tilde{\gamma_\mu }$
are transpose matrix.
 Rearranging function $\widetilde{\psi^{*}}$
and introducing new bispinor $\overline{\widetilde{\psi}}=
\widetilde{\psi^{*}}\gamma_0$, we can obtain \be\label{bar}
\overline{\widetilde{\psi}}\left(\gamma p+m_1 -\gamma_5
m_2\right)=0.\ee The operator $p$ is assumed here acts on the
 function, standing on the left of it.
Using (\ref{et1}) we can write equation (\ref{ModDir}),
(\ref{bar}) in the following form \be \label{1} \left(p\gamma - m
\eta\right)\widetilde{\psi}=0\ee \be\label{2}
\overline{\widetilde{\psi}} \left (p\gamma + m \eta^{-1}
 \right)=0\ee

 Multiplying (\ref{1}) on the left of the
 $\bar{\widetilde{\psi}}
e^{-\vartheta\gamma_5} $ and the equation (\ref{2}) on the right
of the $e^{\vartheta\gamma_5}\widetilde{\psi}$ and summing up the
resulting expressions, one can obtain \be
\bar{\widetilde{\psi}}e^{-\vartheta\gamma_5/2}\gamma_{\mu}
e^{\vartheta\gamma_5/2}(p_{\mu} \widetilde{\psi}) +
(p_{\mu}\widetilde{\bar{\psi}})e^{-\vartheta\gamma_5
/2}\gamma_{\mu}e^{\vartheta\gamma_5/2}\widetilde{\psi}=p_\mu\left(\bar{\widetilde{\psi}}e^{-\vartheta\gamma_5
/2}\gamma_{\mu}e^{\vartheta\gamma_5/2}\widetilde{\psi}\right)=0
\ee Here brackets indicate  which of the  function are  subjected
to  the action of  the operator $p_{\mu}$. The obtained equation
has the
 form  of the continuity equation

 \be \label{j}\partial_\mu j_\mu =0,
 \ee
 where \be\label{21} j_\mu = \widetilde{\bar{\psi}}e^{-\vartheta\gamma_5
/2}\gamma_\mu
 e^{\vartheta\gamma_5}\widetilde{\psi} = \left(\widetilde{\psi^{*}} e^{\vartheta\gamma_5}\widetilde{\psi}, \widetilde{\psi^{*}}\gamma_0\vec{\gamma }
 e^{\vartheta\gamma_5}\widetilde{\psi} \right)\ee

  Thus here the value of $j_\mu $ is a 4-vector of current density of
  particles in the model with $\gamma_5$-mass extension.
   It is very important that its temporal component \be\label{j_0} j_0=\widetilde{\psi}^{*}e^{\vartheta\gamma_5}\widetilde{\psi} \ee
     does not change in time (see (\ref{j}) and positively defined.
   It is easy to see from the following procedure.
Let us construct  the  norm of any state for considered model for
arbitrary vector, taking into account the weight operator $\eta$
(\ref{eta}):
$$
\widetilde{\psi}= \left( \begin{array}{cc}
 x+i y&{} \\
u+iv&{}\\
z+iw&{}\\
t+ip&{}
\end{array}\right).
$$
Using (\ref{chsh}), (\ref{j_0}),  in a result we have

$$
 \widetilde{\psi^{*}}\eta=\left(
\frac{m_1+m_2}{m}(x-iy),\frac{m_1+m_2}{m}(u-iv),
\frac{m_1-m_2}{m}(z-iw), \frac{m_1-m_2}{m}(t-ip)\right).
$$
Then \be \label{Psi} \langle{}
\widetilde{\psi^{*}}\eta|\widetilde{\psi}\rangle=\frac{m_1+m_2}{m}(x^2+y^2)+\frac{m_1+m_2}{m}(u^2+v^2)+
\frac{m_1-m_2}{m}(z^2+w^2)+\frac{m_1-m_2}{m}(t^2+p^2) \ee is
explicitly non negative, because  $m_1\geq m_2$ in the area of
unbroken ${\cal PT}$-symmetry (\ref{e210}).

With the help of (\ref{1}),(\ref{2}) and properties commutation of
$\gamma$-matrix one can obtain that components of new bispinor
amplitudes  must satisfy the following system of algebraic
equations:

\be\label{u} \left(\gamma{p}-m
e^{\gamma_5\vartheta}\right)\widetilde{u}=0; \ee

\be\label{u1} \overline{\widetilde{u}} \left(\gamma{p}-m
e^{-\vartheta\gamma_5}\right)=0, \ee where
$\overline{\widetilde{u}}={\widetilde{u}}^{*}\gamma_0$.

 According  to (\ref{et1}),(\ref{H0}) one can  write bispinor amplitudes
 in the form

\be\label{u2} \widetilde{u}=\sqrt{2m}\left(%
\begin{array}{c}
  A_1 w \\
  A_2 w \\
\end{array}%
\right); \ee

\be \label{u2}\overline{\widetilde{u}}=\sqrt{2m}\left(%
\begin{array}{cc}
 A_1 w^{*}, & - A_2 w^{*} \\
\end{array}%
\right),\ee where the notations are used:
$$
A_1=\cosh\frac{\vartheta}{2}\cosh\frac{\beta}{2}
+\sinh\frac{\vartheta}{2}\sinh\frac{\beta}{2}(\textbf{n}
\overrightarrow{{\sigma}}) ;
$$
$$
A_2= \sinh\frac{\vartheta}{2}\cosh\frac{\beta}{2}
+\cosh\frac{\vartheta}{2}\sinh\frac{\beta}{2}(\textbf{n}
\overrightarrow{{\sigma}}).
$$
In addition, we have  relations (\ref{eta}),(\ref{chsh}) and the
parameters
 $\cosh\beta=\emph{\emph{\emph{E}}}/m$,
$\sinh\beta=p/m$. And also
   $w $ - two-component spinor, satisfying the normalization condition
$$
   w^{*} w =1.
$$
Besides need to note that $\overrightarrow{\sigma}$ are ordinary
$2\times 2$-Pauli matrices and $\textbf{n}=\textbf{p}/p $ - a unit
vector in the direction of the momentum.

The explicit form of these spinors can be found using the
condition that spiral states correspond to the plane wave in which
spinors $w$ is a eigenfunctions of the operator
$(\overrightarrow{\sigma }{\textbf n})$
$$
           \overrightarrow{\sigma}{\textbf n}w^{\zeta}=\zeta w^{\zeta}.
$$
Therefore we get
$$
   w^{1} = \left(%
\begin{array}{c}
  e^{-i\varphi/2}\cos\theta/2 \\
  e^{i\varphi/2}\sin\theta/2 \\
\end{array}%
\right),\,\,\,\,\,                       w^{-1} = \left(%
\begin{array}{c}
  -e^{-i\varphi/2}\sin\theta/2 \\
  e^{i\varphi/2}\cos\theta/2 \\
\end{array}%
\right),
$$
where $\theta$ and $\varphi$ - polar and azimuthal angles that
determine the direction \textbf{n} concerning  to the axes
$x_1,x_2,x_3$.

It is easy to verify by the direct multiplication that
$$
\overline{\widetilde{u}}\widetilde{u} = 2 m.
$$
This result however in advance obvious, because there are the
connection between bispinor amplitudes of modified equations
$\overline{\widetilde{u}},\widetilde{u}$ and  corresponding
solutions of the ordinary Dirac equations:
 $$ \widetilde{u}= e^{-
\gamma_5\vartheta/2}u$$ $$ \overline{\widetilde{u}}= \overline{u}
e^{\gamma_5\vartheta/2}.
 $$
 Taking into account that  the  Dirac bispinor amplitudes as usually Ref.\cite{TKR}
are normalized by invariant condition $\overline{u} u =2m$. Hence
we have \be\label{u1u} \overline{\widetilde{u}}\widetilde{u}
=\overline{u} u = 2m. \ee

By using (\ref{u}),(\ref{u1}) and(\ref{u1u}) we can also obtain
$$
\overline{\widetilde{u}}e^{-\gamma_5 \vartheta}\gamma_\mu
\widetilde{u}=2 p_\mu.
$$
Taking into account (\ref{psi}) and (\ref{21}),(\ref{j_0}) one can
easily find

$$
    j_{\mu}= \frac{1}{2E} \tilde{\bar{u}}\gamma_{\mu}\eta \tilde{u}  =\{ 1, \textbf{p}/E  \},
$$
whence it follows that the operator $\eta=e^{\gamma_5 \vartheta}$
in full compliance with Mostafazadeh's  result (see, for example
Ref. \cite{ali},\cite{alir}) induces the inner product
  $$
      \widetilde{\psi}^{*}\eta  \widetilde{\psi}= 1,
  $$
for $\widetilde{\psi}\neq 0$.

\section{Dirac modified models with $\gamma_5$-massive
contributions in the external homogenies magnetic field}

As it is known the wave Dirac equations provide a basis for
relativistic quantum  mechanics and quantum electrodynamics of
spinor particles in  external electromagnetic fields. Exact
solutions of relativistic wave  equation are referred to as
one-particle wave functions which
 allow the development of the approach known as the Furry picture.
This method incorporates study the interactions with the external
field exactly, regardless of the field  intensity(see
Ref.\cite{TKR}). Without the knowledge of exact solutions there is
no regular methods of describing such interactions with an
arbitrariness field  explicitly. The physically most important
exact solutions of the ordinary Dirac equations are: an electron
in a Coulomb field, in a uniform magnetic field and in  the field
of a plane wave. In this connection, is of interest  the
investigating  of non-Hermitian Dirac models which describe an
alternative formulation of relativistic quantum mechanics where
the Furry picture also may be realized.

Consider a uniform magnetic field $\textbf{H}=(0,0,H)$ directed
along the $x_3$ axis ($H > 0$). The electromagnetic potentials are
chosen in the  gage Ref.\cite{TKR} \be
   A_0=0,\,\,A_1=0,\,\,A_2=H x_1\,\,  A_3=0.
\ee We can write the modified Dirac equations in the form
\be\label{cal P}
  \left( \gamma_\mu {\cal P^\mu} -m e^{\vartheta
  \gamma_5}\right)\widetilde{\Psi}=0,
\ee were ${\cal P^\mu} = i\partial_{\mu} -e A_\mu $ ; $e=-|e|$ and
$\gamma$- matrixes still chosen in the standard representation. In
the field under consideration, the operators ${\cal P}_0,\,{\cal
P}_2 $ and ${\cal P}_3$ are mutually commuting integrals  of
motion $[{\cal D},{\cal P}_0]=0$,$[{\cal D},{\cal P}_2]=0$,
$[{\cal D},{\cal P}_3]=0$, where ${\cal D}=( \gamma_\mu {\cal
P}^{\mu} - m e^{\vartheta
  \gamma_5}) $.

Let present the function $\widetilde{\Psi }$ in the form
$$
        \widetilde{ \Psi} = \left(%
\begin{array}{c}
  \psi_1 \\
  \psi_2 \\
  \psi_3 \\
  \psi_4 \\
\end{array}%
\right)e^{-i E t}
$$
and use Hamilton's form of Dirac  equations \be
\label{H2}H\widetilde{\psi} = E\widetilde{\psi} \ee where
$$
H=(\overrightarrow{\alpha} \textbf{{\cal P}}) +\beta m_1
+\beta\gamma_5 m_2.
$$
It is useful to introduce the change of variables Ref.\cite{TKR}
$$
  \psi_i(x_1,x_2,x_3)= e^{ip_2 x_2+ip_3 x_3}\Phi_i(x_1),
$$ where $i=1,2,3,4.$
 we can obtain the following system of equations:
\be\label{sist1}(E\mp m_1)\Phi_{1,3}+iR_2 \Phi_{4,2} -(p_3 \mp
m_2) \Phi_{3,1} =0;\ee where $ R_2=\left[\frac{\partial}{\partial
x_1} +(p_2+{e H})\right];$ \be\label{sist}
 (E\mp
m_1)\Phi_{2,4}+iR_1\Phi_{3,1}+(p_3 \pm m_2) \Phi_{4,2}=0. \ee Here
$R_1=\left[\frac{\partial}{\partial x_1} -(p_2+{e H})\right]$ and
top mark  relates to the components of the wave function with the
first index, and the lower - to the components with the second
index

Next convenient to go to the dimensionless variable \be\rho =
\sqrt{\gamma}x_1 +p_2/\sqrt{\gamma},  \ee where $\gamma=|e| H$,
and equations (\ref{sist1}),(\ref{sist}) take the form
\be\label{PH1}
 (E\mp m_1)\Phi_{1,3} + i
 \sqrt{\gamma}\left(\frac{d}{d\rho}+\rho\right)\Phi_{4,2}-(p_3 \mp
 m_2)\Phi_{3,1} =0;
\ee \be\label{PH2} (E\mp m_1)\Phi_{2,4} + i
 \sqrt{\gamma}\left(\frac{d}{d\rho}-\rho\right)\Phi_{3,1}+(p_3 \pm
 m_2)\Phi_{4,2} =0.
\ee General solution of this system can be represented in the form
of the Hermite functions
$$
u_n(\rho)=\left(\frac{\gamma^{1/2}}{2^n n! \pi^{1/2}}
\right)e^{-\rho^2 /2}H_n(\rho),
$$
where $H_n(x)$ is standardizing the Hermite polynomials:
$$
  {\mathit{H}}_{n}(x)=(-1)^n
e^{x^2/2}\frac{d^n}{dx^n}e^{-x^2/2},
$$
and $n=0,1,2..$. In should be noted that Hermite function are
satisfied to the recurrent relations: \be\label{u11}
\left(\frac{d}{d\rho}+\rho\right)u_n=(2n)^{1/2}u_{n-1}; \ee
\be\label{u22}
\left(\frac{d}{d\rho}-\rho\right)u_{n-1}=-(2n)^{1/2}u_{n}.\ee It
is easy to see  from (\ref{u11}),(\ref{u22}) that
$$
   \left(\frac{d}{d\rho}-\rho \right)\left(\frac{d}{d\rho}+\rho
   \right)u_n = -2n u_n
$$
and hence ( see, for example Ref.\cite{TKR} ) \be\label{R1R2}
   R_1 R_2 =-2\gamma n.
\ee

Substituting next in (\ref{PH1}),(\ref{PH2})
$$
\Phi=\left(%
\begin{array}{c}
  C_1 u_{n-1}(\rho)\\
  iC_2 u_n(\rho)\\
  C_3 u_{n-1}(\rho)\\
  iC_4 u_{n}(\rho)\\
\end{array}%
\right),
$$
one can fined that coefficients $C_i\,(i=1,2,3,4)$ are determined
by algebraic equations
$$
(E\mp m_1)C_{1,3}-(2\gamma n)^{1/2} C_{4,2} -(p_3 \mp m_2)C_{3,1}
=0;
$$
$$
(E\mp m_1)C_{2,4}-(2\gamma n)^{1/2} C_{3,1}+(p_3 \pm m_2)
C_{4,2}=0.
$$
The equality to zero of the determinant of this system leads to a
spectrum of energy of the non-Hermitian Hamiltonian in the form
\be\label{40}
    E=\pm\sqrt{{m_1}^2-{m_2}^2 + 2\gamma n +{p_3}^2},
\ee where $n=0,1,2..$, and with take into account
$m^2={m_1}^2-{m_2}^2$, we can see the result, which also
(see(\ref{E})) coincides with the eigenvalues of  Hermitian
Hamiltonian
 Ref.\cite{TKR}.

The coefficients $C_i$ may be determined if one uses operator of
polarization in the form of the third component of the
polarization tensor in the direction of the magnetic field
\be\label{muH}
          \mu_3=m_1\sigma_3 + \rho_2[\vec{\sigma}\vec{{\cal
          P}}]_3
\ee where matrices $$ \sigma_3= \left(%
\begin{array}{cc}
  I & 0 \\
  0 & -I \\
\end{array}%
\right); \,\,\,\,\,              \rho_2 = \left(%
                                      \begin{array}{cc}
                                        0 & -iI \\
                                     iI & 0 \\
                                         \end{array}%
                                       \right).
$$

It is easy to see, that bispinor $C$ can be written as

         \be\label{PsiH1} \left(%
\begin{array}{c}
  C_1 \\
  C_2 \\
  C_3 \\
  C_4 \\
\end{array}%
\right)=\frac{1}{2\sqrt{2}}\left(%
\begin{array}{c}
  \cosh(\vartheta/2) \Phi_1+\sinh(\vartheta/2) \Phi_3 \\
  \cosh(\vartheta/2) \Phi_2+\sinh(\vartheta/2) \Phi_4 \\
  \sinh(\vartheta/2) \Phi_1 +\cosh(\vartheta/2) \Phi_3\\
  \sinh(\vartheta/2) \Phi_2 +\cosh(\vartheta/2) \Phi_4, \\
\end{array}%
\right),\ee where
$$
\Phi_1=\sqrt{1+\zeta m/p_\bot}\sin(\pi/4+\lambda/2)
$$
$$
\Phi_2=\zeta\sqrt{1-\zeta m/p_\bot}\sin(\pi/4-\lambda/2)
$$
$$
\Phi_3=\zeta\sqrt{1+\zeta m/p_\bot}\sin(\pi/4-\lambda/2)
$$
$$
\Phi_4=\sqrt{1-\zeta m/p_\bot}\sin(\pi/4+\lambda/2).
$$
Here $\mu_3 \psi =\zeta k\psi$, $k=\sqrt{{p_\bot}^2 + m^2}$ and
$\zeta=\pm 1 $ that is corresponding to the orientation of the
fermion spin: along $(+1)$ or opposite $(-1)$ to the magnetic
field, and parameter $ \lambda$ obey to the condition
$\cos{\lambda}=p_3/E.$ The functions $sinh(\vartheta/2)$ and
$cosh(\vartheta/2)$ are defined by relations (\ref{chsh}).

\section{ Non-Hermitian modified Dirac-Pauli  model in the magnetic
field}

In this section, we will touch upon a question of describing the
motion of Dirac particles, if their own magnetic moment is
different from the Bohr magneton. As it was shown by Schwinger
Ref.\cite{Sc}, that  the Dirac equation of particles in the
external electromagnetic field $A^{ext}$ taking into account the
radiative corrections may be represented in the form \be\label{A}
\left({\cal P}\gamma -m\right)\Psi(x)-\int{\cal
M}(x,y|A^{ext})\Psi(y)dy=0, \ee where ${\cal M}(x,y|A^{ext})$ is
the mass operator of fermion in external  field. From equation
(\ref{A}) by means of expansion of the mass operator in series
according to  $ eA^{ext}$ with precision not over then linear
field terms  one can obtain the modified equation( see, for
example, Ref.\cite{TKR}). This equation preserves the relativistic
covariance and consistent with the phenomenological equation of
Pauli obtained in his early papers.

Now let us consider the model of massive fermions with
$\gamma_5$-extension of mass $m\rightarrow m_1+\gamma_5 m_2$
taking into account the interaction of their charges and AMM with
the electromagnetic field $F_{\mu\nu}$:

\be\label{Delta} \left( \gamma^\mu {\cal P}_\mu -
 m_1 -\gamma_5 m_2 -\frac{\Delta\mu}{2}\sigma^{\mu \nu}F_{\mu\nu}\right)\widetilde{\Psi}(x)=0,\ee
where $\Delta\mu = (\mu-\mu_0)= \mu_0(g-2)/2$. Here $\mu$ -
magnetic moment of a fermion, $g$ - fermion gyromagnetic factor,
$\mu_0=|e|/2m$ - the Bohr magneton,
$\sigma^{\mu\nu}=i/2(\gamma^\mu \gamma^\nu-\gamma^\nu
\gamma^\mu)$. Thus phenomenological constant $\Delta\mu $, which
was introduced by Pauli,  is part of the equation and gets the
interpretation with the point of view quantum field theory.

The Hamiltonian form of (\ref{Delta}) in the homogenies magnetic
field is the following \be i\frac{\partial}{\partial t}
\widetilde{\Psi}(r,t)=H_{\Delta \mu}\widetilde{\Psi}(r,t),\ee
where \be\label{Delta1} H_{\Delta\mu} = \vec{\alpha}\vec{{\cal P}}
+ \beta(m_1 + \gamma_5 m_2) +
\Delta\mu\beta(\vec{\sigma}\textbf{H}).\ee Given the quantum
electrodynamic contribution in AMM of an electron with accuracy up
to $e^2$ order we have $\Delta\mu=\frac{\alpha}{2\pi}\mu_0 $,
where $\alpha = e^2 =1/137$ - the fine-structure constant and we
still believe that the potential of an external field satisfies to
the free Maxwell equations.

It should be noted that now the operator projection of the fermion
spin at the direction of  its movement - $\overrightarrow{ \sigma}
\overrightarrow{{\cal P }} $ is not commute with the Hamiltonian
(\ref{Delta1}) and hence it is not the integral of motion. The
operator, which is commuting with this Hamiltonian  remains
$\mu_3$ (see (\ref{muH})). Subjecting the wave function
$\widetilde{ \psi }$ to requirement to be eigenfunction of the
operator polarization (\ref{muH}) and Hamilton operator
(\ref{Delta1})we can obtain: \be\label{Pi}
\mu_3\psi = \zeta k\psi, \,\,\, \mu_3=\left(%
\begin{array}{cccc}
  m_1 & 0 & 0 & {\cal P}_1-i{\cal P}_2 \\
  0 & -m_1 & -{\cal P}_1-i{\cal P}_2 & 0 \\
  0 & -{\cal P}_1+i{\cal P}_2 & m_1 & 0 \\
  {\cal P}_1+i{\cal P}_2 & 0 & 0 & -m_1 \\
\end{array}%
\right), \ee where $\zeta=\pm 1$ are characterized the projection
of fermion spin at the direction of the magnetic field.
 $$H_{\Delta\mu}\widetilde{\psi}=E\widetilde{\psi},$$
  \be\label{Hmu} H_{\Delta\mu}=\left(%
\begin{array}{cccc}
  m_1+H\Delta\mu & 0 & {\cal P}_3 -m_2& {\cal P}_1-i{\cal P}_2 \\
  0 & m_1-H\Delta\mu & {\cal P}_1+i{\cal P}_2  & -m_2-{\cal P}_3\\
  m_2+{\cal P}_3 & {\cal P}_1-i{\cal P}_2 & -m_1-H\Delta\mu & 0 \\
  {\cal P}_1+i{\cal P}_2 & m_2-{\cal P}_3 & 0 & H\Delta\mu-m_1 \\
\end{array}%
\right). \ee

A feature of the model with $ \gamma_5 $-mass contribution is that
it may contain another any restrictions of mass parameters in
addition to (\ref{e210}). Indeed while that for the physical mass
$m$ one may be constructed by infinite number combinations of $
m_1 $ and $ m_2 $, satisfying to (\ref{012}), however besides it
need take into account the rules of conformity of this parameters
in the Hermitian limit. Without this the developing of
Non-Hermitian models may not be adequate. With this purpose one
can determine an additional variable, which will depend on $m_1$,
$m_2$ and which would put an upper bound on the mass spectrum of
particles. In particular, the simple a linear scheme may be
constructed easy if one takes the obvious restriction for the mass
spectrum of the fermions by using \be \label{m1111} m \leq M1, \ee
where $M1 $ is the auxiliary parameter which is equal to the value
of mass parameter $ m_1$($M_1=m_1$).  Using this approach we can
describe in  principle  the whole spectrum of fermions when $m
\leq M_1$ by means of defining appropriate values $ m_2 $.

 According to
(\ref{012}) one can obtain the expression \be \label{M11}  m =
M_1\sqrt { 1 - {m_2} ^ 2/{M_1}^2 }. \ee  With the help (\ref
{M11}) we can see,  that when the mass $ m_2 $ is increased,  the
values of the physical mass tends  to zero (see Fig.1).  The
equality of parameters $ m_2 = m_1 $ corresponds to the case of
massless fermions. But it should be noted that in this linear
model the Hermitian limit $ m_2 \rightarrow 0 $ may be  reached
only in the case of the the particles with maximal mass $ m = M1
$. At the same time, \emph{the Hermitian limit is absent for  all
other  mass values}.

 Thus the procedure of limitations of  the
physical mass spectrum by the inequality $ m \leq M1 $ has the
essential drawback since in this frame is not possible to describe
all ordinary fermions, respecting to the  Principle of Conformity,
except, $ m = M1 $.

These considerations make search in the frame of mass restriction
of $m\leq m_1$ the existence of more complicated non-linear
dependence of limiting mass value \be\label{mM} m \leq
M(m_1,m_2),\ee which meets the requirements of the  Principle of
Conformity:

i)The Dirac limit must exist for all ordinary fermions for  which
 the condition(\ref{mM}) is satisfied.

ii)In Hermitian limit  the parameter $m_1$ must coincide with any
physical mass $m$.

 The fulfillment of these conditions lets to find the most
appropriate scheme  restriction of  the mass fermions,  for  which
exist  the consistency with the ordinary Dirac theory.
 Possibly explicit expression for $M(m_1,m_2)$, may be obtained
from the \textbf{simple mathematical theorem:} \emph{the
arithmetical average of two non-negative real numbers $a$ and $b$
which always is greater than or equal to the geometrical mean of
the same numbers.}
 Really, let $a=m^2$ and $b={m_2}^2$ then using
$$ \frac{{m}^2+{m_2}^2}{2}\geq\sqrt{m^2\cdot {m_2}^2}$$ and
substitution (\ref{012}), we can get the inequality \be\label{mM1}
m\leq {m_1}^2/2m_2 = M(m_1,m_2).\ee Values of $M$ is now defined
by two parameters $m_1,m_2$. And in the Hermitian limit, when $m_2
\rightarrow 0$ the value of the maximal mass $M$ tends to
infinity. It is very important that in this limit the restriction
of mass values completely disappear. In such a way one can
demonstrate a natural transition from the Modified Model to the
Standard Model (SM), which contains any values of the physical
mass $m$.

Using (\ref{012}) and expression (\ref{mM1}) we can also obtain
the system of two equations \be\label{sys}\left\{
\begin{array}{c}
  m={m_1}^2-{m_2}^2 \bigskip\\
  M={{m_1}^2}/{2 m_2} \\
\end{array}\right.
\ee  The solution of this system  relative to the parameters $m_1$
and $m_2$ may be represented in the form

\be\label{m11111} {m_1}^{\mp} =\sqrt{2}M
\sqrt{1\mp\sqrt{1-m^2/{M}^2}}; \ee

\be \label{m22222}{m_2}^\mp =  M\left(1\mp \sqrt{1-m^2/{M}^2}
\right). \ee

It is easy to verify that the obtained values of the mass
parameters satisfy the conditions (\ref{012}) and (\ref{e210})
regardless of which the sign will be chosen. Besides  formulas
(\ref{m11111}),(\ref{m22222}) in the case of the upper sign are
agreed with conditions $m_2\rightarrow 0$ and $m_1\rightarrow m$
when $M \rightarrow \infty,$ i.e. when a the  Hermitian  limit is
exist.  However, if one choose a lower sign (i.e. for the
${m_1}^+$ and ${m_2}^+$)  such limit is absent. Thus we can see
that the nonlinear scheme of mass restrictions (see (\ref{mM1}))
additionally contains the solutions satisfying to the some new
particles. However this solutions should be considered only as an
indication of the principal possibility of the existence of such
particles. In this case, as follows from
(\ref{m11111}),(\ref{m22222}) for each ordinary particles may be
exist some new partners, possessing the same mass and a number of
another unique properties.\footnote{As the exotic particles do not
agree in the "flat limit"
 with the ordinary Dirac expressions then one can
assume that in this case we deal with a description of some new
particles, properties of which have not yet been studied. This
fact for the first time has been fixed by V.G.Kadyshevsky in his
early works in the geometric approach to the development of the
quantum field  theory with a fundamental mass" Ref.\cite{Kad1} in
curved de-Sitter momentum space. Besides in
Ref.\cite{KMRS},\cite{Max} it was noted that the most intriguing
prediction of the new approach is the possible existence of exotic
fermions with no analogues in the SM, which may be candidates for
constituents of dark matter .}

\begin{figure}[h]
\vspace{-0.2cm} \centering
\includegraphics[angle=0, scale=0.5]{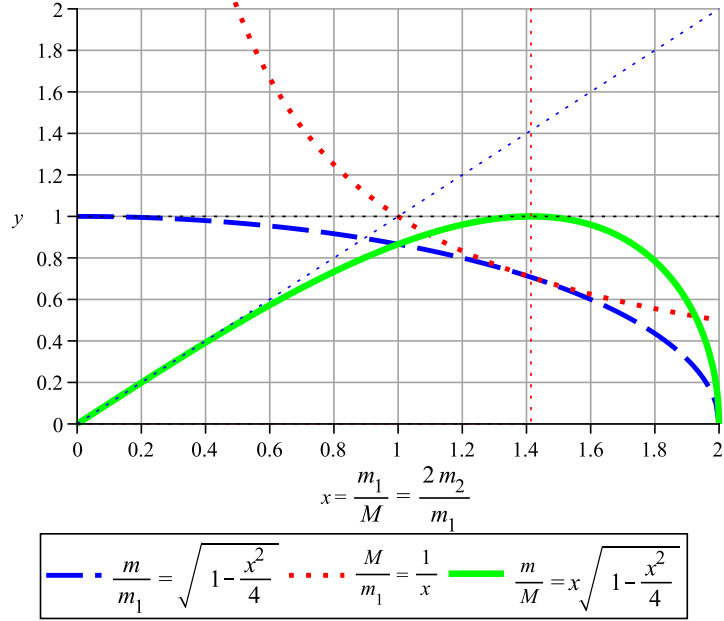}
\caption{Dependence of $m/M,  M/m_1$, and  $m/m_1 $ on the
parameter $x=m_1/M=2m_2/m_1$} \vspace{-0.1cm}\label{Fig.1-1}
\end{figure}

Let's consider the  "normalized" parameter of the modified model
with the maximal mass $M$: \be \label{x} x=\frac{m_1}{M}=\frac{2
m_2}{m_1}\ee and using (\ref{M11}) we can obtain \be\label{y}
\frac{m_2}{M}=\frac{x^2}{2}\ee \be \frac{m}{M}=x\sqrt{1-x^2/4}\ee

At Fig.1 one can see the dependence of the normalized parameters
$m/m_1$, $M/m_1$ and $m/M$ on the relative parameter
$x=m_1/M=2m_2/m_1$. In particular, the maximum value of the
particle mass $ m = M $ is achieved at the ratio of the subsidiary
masses is equal to $m_2 = m_1/ \sqrt{2}$. Till to this value for
each mass of ordinary particles, one can find the parameters $m_1$
and $m_2$, for which a limit transition to regular theory Dirac
exist. Further increasing of $m_2$, leads to the descending branch
of the $m/M$, where the   Dirac limit  is  not exist  and at the
point $m_2=m_1$ the value of $m$ is equal to zero. Thus, it is the
region $m_1 > \sqrt{2}m_2$ ($m_2 > M$ ) corresponds to the
description of the "exotic particles", for which there is not
transition to Hermitian limit.

Thus, in the frame of the known  inequality  (\ref{e210})  we can
see  three specific sectors of unbroken $\cal PT$-symmetry of the
 non-Hermitian  Hamiltonian (\ref{H}) in the plane
$\nu_1=m_1/M, \nu_2=m_2/M$.  Thus the plane $\nu_1,\nu_2$ may be
divided by the three groups of the inequalities:

$$I.\,\,\,\,\,\,\,\,\,\,\,\,\,\,\,\,\,\,\, \nu_1/\sqrt{2} \leq \nu_2 \leq\nu_1,$$
$$II.\,\,\,\,-\nu_1/\sqrt{2}< \nu_2 < \nu_1/\sqrt{2},$$
$$III. \,\,\,\,\,\,\,\, -\nu_1 \leq \nu_2\leq-\nu_1/\sqrt{2},$$

It is very important that only the region $II.$ corresponds to the
description of ordinary particles, while  $I.$ and$. III.$ define
the description of some as yet unknown particles. This conclusion
is not trivial, because in contrast to the geometric approach,
where the emergence of new unusual properties of particles
associated with the presence in the theory a new degree of freedom
(sign of the fifth component of the momentum
$\varepsilon=p_5/|p_5|$ \cite{KMRS}), in the case of a simple
extension of the free Dirac equation due to the additional
$\gamma_5$-mass term, the satisfactory explanation of this fact is
not there yet.


Then we can  establish the limits of change of parameters. As it
follows from the (\ref{m11111}),(\ref{m22222})  the limits of
variation of parameters $m_1$ and $m_2$ are the following:
\be\label{limit} m\leq m_1\leq 2 M; -2M \leq m_2 \leq 2M.\ee  In
the areas of change of these parameters  a point  exists in which
 we have \be\label{m111}  m_1=\sqrt{2}M;\,\, m_2=M.\ee  In this point the
physical mass $m$  reaches its maximum value $m=M$ (see Fig.1)
that corresponds to the mass of  the "maximon" Ref.\cite{Kad1}.

Performing calculations  here in many ways reminiscent of similar
calculations carried out in the previous section. In a result  of
modified Dirac-Pauli equation one can also find    \emph{the exact
solution for energy spectrum}:
 \be\label{E61} E(\zeta,p_3,2\gamma
n,H)=\pm\sqrt{{p_3}^2-{m_2}^2+\left[\sqrt{{m_1}^2+2\gamma
n}+\zeta\Delta\mu H \right]^2} \ee and for eigenvalues of the
operator polarization $\mu_3$ we can write in the form \be
k=\sqrt{{m_1}^2 +2\gamma n}. \ee

It is easy to see that in the case $ \Delta\mu =0$ from
(\ref{E61}) one can obtain the expression (\ref{40}).  Besides it
should be emphasized that the expression analogical to
(\ref{E61}), in the frame of ordinary Dirac-Pauli approach one can
obtain putting $m_2=0$ and $m_1=m$: \be\label{DPA}
E(\zeta,p_3,2\gamma
n,H)=\pm\sqrt{{p_3}^2+\left[\sqrt{{m}^2+2\gamma n}+\zeta\Delta\mu
H \right]^2}.
 \ee

Note that in the paper Ref.\cite{TBZ} was previously obtained
result analogical to  (\ref{DPA}) by means of using the ordinary
Dirac-Pauli approach. Direct comparison of modified formula
(\ref{E61}) in the Hermitian limit with the  result Ref.\cite{TBZ}
shows their coincidence. It is easy to see that the expression
(\ref{E61}) contains dependence on  parameters $m_1$ and $m_2$
separately, which are not combined into \emph{a mass of particles}
$m=\sqrt{{m_1}^2-{m_2}^2}$ that essentially differs from the
examples which were considered in previous  sections
\textbf{2,\,3}.

 Thus, in contrast to (\ref{E}) and (\ref{40}) here the calculation of
interaction AMM of fermions with the magnetic field allow to put
the question about the possibility of experimental studies of the
effects of $\gamma_5$-extensions of a fermion mass. In particular
if to suggest that $m_2=0$ and hence $m_1 =m$, we obtain as noted
earlier, the Hermitian limit. But taking into account the
expressions (\ref{m11111}) and (\ref{m22222}) we obtain that the
energetic spectrum (\ref{E61}) is expressed through the fermion
mass $m$ and the value of the maximal mass $M$. Thus, taking into
account that the interaction AMM with magnetic field removes the
degeneracy on spin variable, we can obtain the energy of the
ground state ($\zeta=-1$) in the form
 \be\label{E1}
E(-1,0,0,H,x)=m\sqrt{-\left({\frac{1\mp\sqrt{1-x^2}}{x}}\right)^2+
\left(\frac{\sqrt{2}\sqrt{1\mp\sqrt{1-x^2}}}{x}-\frac{\Delta\mu
H}{m} \right)^2}, \ee where $x=m/M$ and the upper sign corresponds
to the ordinary particle and the lower sign defines their "exotic"
partners.

Through decomposition of functions ${}^{-}m_1$ and ${}^{-}m_2$ we
can obtain

\be\label{sys}{}^{-}m_1/m =\left\{
\begin{array}{c}
  1+\frac{x^2}{8}+\frac{7 x^4}{128},\,x\ll 1 \bigskip\\
  \frac{\sqrt{2}}{x}\qquad \qquad x\rightarrow 1 \\
\end{array}\right.\,\,\,\,\,
{}^{-}m_2/m =\left\{
\begin{array}{c}
   \frac{x}{2}+\frac{x^3}{8}+\frac{x^5}{16},\,\,\,\,\,\,\,\,x\ll 1\bigskip\\
  \frac{1}{x}\qquad\qquad\qquad x\rightarrow 1 \\
\end{array}\right.
\ee

Similarly, for ${}^{+}m_1$ and ${}^{+}m_2$ we have

\be\label{sys1}{}^{+}m_1/m =\left\{
\begin{array}{c}
  \frac{2}{x}-\frac{x}{4}-\frac{5 x^3}{64},\,x\ll 1\bigskip \\
  \frac{\sqrt{2}}{x}\qquad \qquad x\rightarrow 1 \\
\end{array}\right.\,\,\,\,\,
{}^{+}m_2/m =\left\{
\begin{array}{c}
   \frac{2}{x}-\frac{x}{2}-\frac{x^3}{8},\,x\ll 1\bigskip\\
  \frac{1}{x}\qquad\qquad\,\, x\rightarrow 1 \\
\end{array}\right.
\ee

\begin{figure}[h]
\vspace{-0.2cm} \smallskip
\includegraphics[angle=0, scale=0.5]{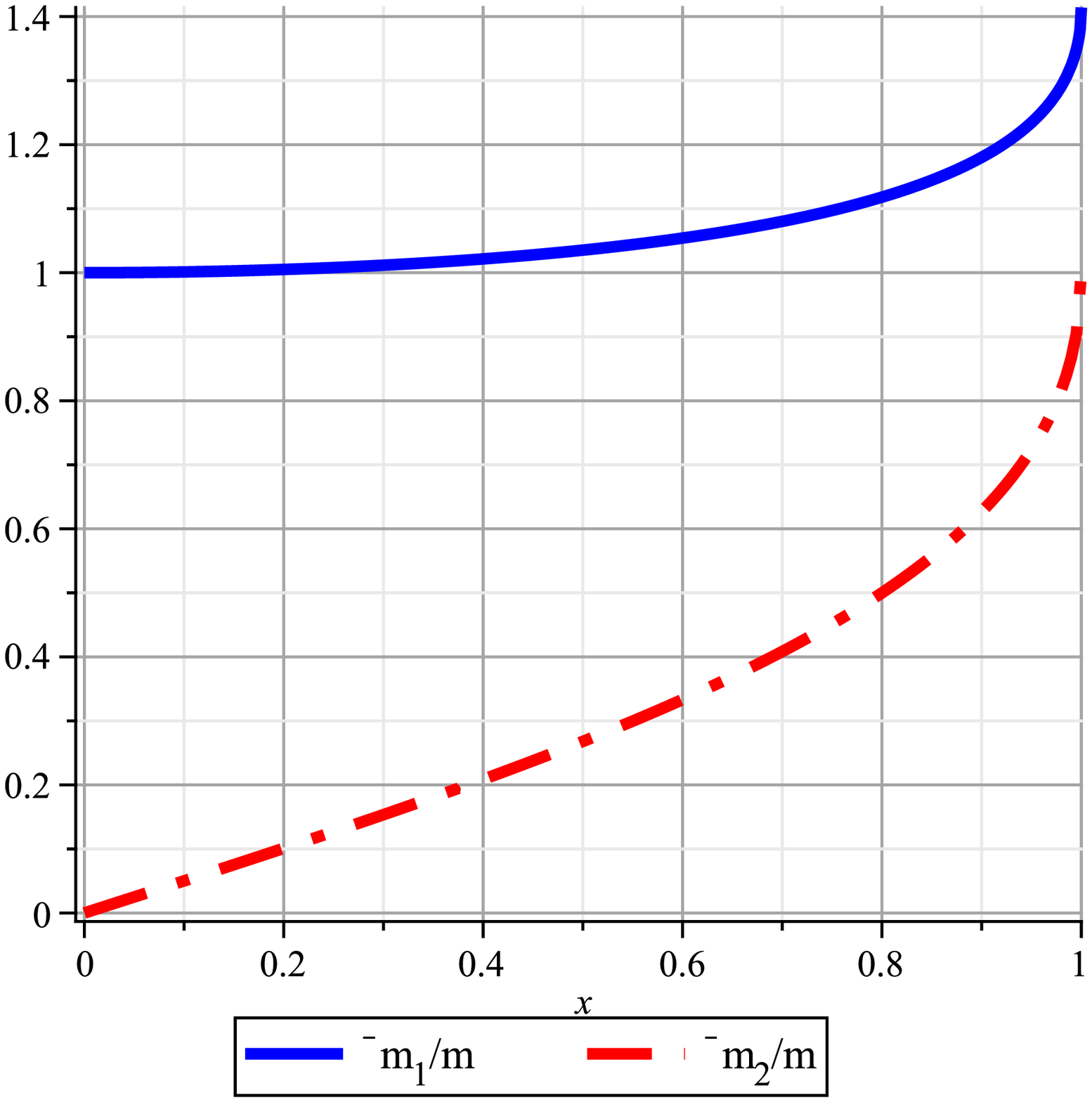}
\caption{The dependence of parameters $^{-}m_1/m, ^{-}m_2/m $ on
the $x=m/M.$} \vspace{-0.1cm}\label{f4}
\end{figure}

\begin{figure}[h]
\vspace{-0.2cm} \smallskip
\includegraphics[angle=0, scale=0.5]{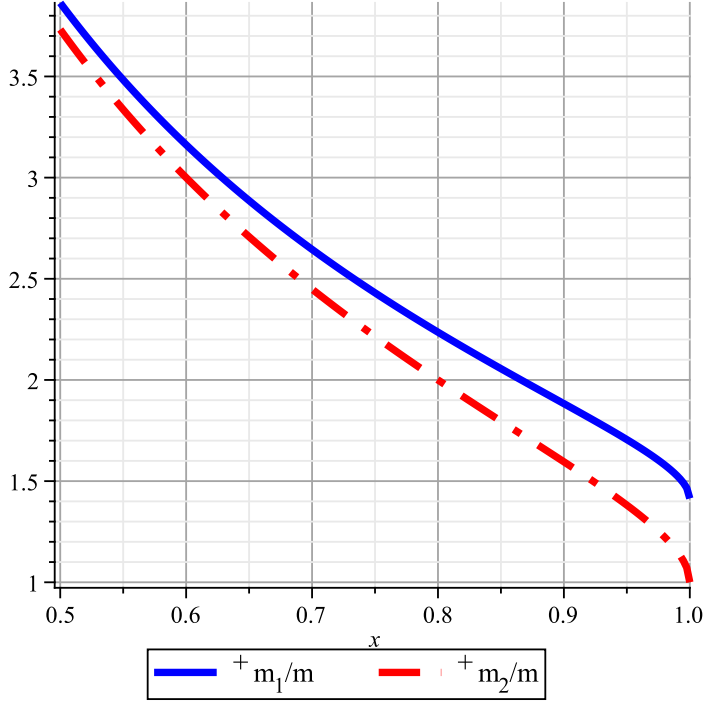}
\caption{The dependence of parameters $^{+}m_1/m, ^{+}m_2/m $ on
the $x=m/M.$} \vspace{-0.1cm}\label{f5}
\end{figure}

Let us now turn to a more detailed consideration of the fermion
energy in the ground state in the external field. As follows from
(\ref{sys}) and (\ref{sys1}) function (\ref{E61}) not trivial
depends on the parameters $x=m/M$ and $H$. For reasons outlined
above, the effect of magnetic field on the energy state of the
ordinary fermion with the small mass $x\ll 1$ (see Fig.2)we can
obtain  in the form

\be\label{E1} E(-1,0,0,H,x)=
m\sqrt{1-\frac{\Delta\mu}{\mu_0}\frac{H}{H_c}(1+x^2/8 + 7 x^4/128)
+ \left(\frac{\Delta\mu H}{m}\right)^2},  \ee where $H_c = m^2/e$.

On the other hand,  for the case of the "exotic" particles in the
similar limit $x\ll1$ the result is significantly different (see
(\ref{sys1}) and Fig.3) \be\label{E2} E(-1,0,0,H,x)= m
\sqrt{1-\frac{\Delta\mu}{\mu_0}\frac{2
 H}{x H_c}+ \left(\frac{\Delta\mu H}{m}\right)^2}. \ee From (\ref{E2})
 one can see that  the field corrections in this case are  substantially
 increased  as  $1/x=M/m \gg 1$.

If the use the limit of the large mass $ x \rightarrow 1$ we get
the matching results as for traditional and exotic particles.
Thus, by combining these results, we can write
 \be\label{E3}
  E(-1,0,0,H,x)= m\sqrt{1-\frac{\Delta\mu}{\mu_0}\frac{\sqrt{2}
 H}{x H_c}+ \left(\frac{\Delta\mu H}{m}\right)^2}.
 \ee

 One can also see from (\ref{sys}),(\ref{sys1}) that the changes of the
parameters ${}^{\mp}m_1 $ and ${}^{\mp}m_2$ occur  by such a way
that in the  point $ x=1$ ($m=M$) the branches of ordinary and
exotic particles are crossed. At Fig.2 and Fig.3  dependencies of
${m}^{\mp}_1/m$ and $m^{\mp}_2/m$ on the parameter $x=m/M$ are
represented and one may clearly see the justice of this fact.

 As the equation (\ref{Delta}) and following from it formulas
(\ref{E61}),(\ref{E1}),(\ref{E2}) would fair for any intensities
magnetic field, it is easy to see with the values in \be H\sim
\frac{\mu_0 }{\Delta\mu}\frac{m}{m_1} H_c,\ee we would obtain
$E_0\sim 0$. Hence, in the intensive magnetic fields accounting of
the vacuum magnetic moment can lead to a substantial change of
borders of the energetic spectrum between the fermion and anti
fermion states. Notice once more that a considerable increasing of
this amendment is connected with
 the possible contributions from so-called exotic
particles.

\section{Summary and
Conclusions.}

In the researches, presented in the previous sections, shown that
the Dirac Hamiltonian of a particle with $\gamma_5$ - dependent
mass term is non-Hermitian, and has the unbroken $\cal PT$ -
symmetry in the area ${m_1}^2\geq {m_2}^2,$ which has three the
subregion. Indeed with the help of the algebraic transformations
we obtain a number of the consequence of the relation (\ref{012}).
In particular there is the restriction of the particle mass in
this model: $m\leq M$, were $M={m_1}^2/2 m_2$. Outside of this
area the $\cal PT$ - symmetry of the modified Dirac Hamiltonians
is broken and the inequality  $m\leq M $ may be considered as a
\emph{new wording of the conditions of unbroken $\cal PT$ -
symmetry. }

In addition, pay attention, that the introduction of limitation of
the mass spectrum, on the basis of a geometric approach to the
development of the modified QFT (see, for example
\cite{KMRS},\cite{Max}), also leads to appearance of non-Hermitian
$\cal PT $-symmetric Hamiltonians in the fermion sector of the
model with the Maximal Mass. On the other hand, it was shown by
us, that non-Hermitian $\cal PT$-symmetric algebraic approach with
$\gamma_5$ - mass term,  may be  considered as a condition of
occurrence of the analogical auxiliary mass parameter of the
model.

 In particular, this applies to the
modified Dirac equation in which was produced  the substitution
$m\rightarrow m_1+ \gamma_5 m_2 $. Into force of the ambiguity of
the definition of the parameter $m$ depending on $m_1, m_2$, the
basic inequality $m_1\geq m_2\geq 0$ contains description of
particles of two species. In the first case, it is
 ordinary particles, when mass parameters are limited by the terms

\be \label{01}     0\leq m_2 \leq m_1/\sqrt{2} . \ee

In the second area we are dealing with so-called «exotic partners»
of ordinary particles, for which is still accomplished
(\ref{e210}), but one can write

\be \label{02} m_1/\sqrt{2} \leq m_2 \leq m_1. \ee

 Intriguing difference  of the second type particles from traditional
 fermions is that they  are described by the different modified Dirac
 equations. So, if in the first case(\ref{01}), the
equation of motion under the transition  $M \rightarrow \infty$
leads to the standard Dirac equation, but  in the second case such
a transition is not there.

Thus, it is shown that the main progress, is obtained by us in the
algebraic way of the construction of the fermion model with
$\gamma_5$-mass term is consists of describing of the new
energetic scale, which is defined by the parameter
$M={m_1}^2/2m_2$. This value on the scale of the masses is a point
of transition from the ordinary particles to exotic. Furthermore,
description of the exotic particles in the algebraic approach are
turned out essentially the same as in the model with a maximal
mass, which was investigated by V.G.Kadyshevsky with colleagues on
the basis of geometrical approach (see, for example
Ref.\cite{Kad1}-\cite{Max}).

We have presented a number of examples of non-Hermitian models
with $\gamma_5$-extension mass in relativistic quantum mechanics
including in presence of external electromagnetic field for which
the Hamiltonian $H$ has a real spectrum. Although the energy
spectra of the fermions in some cases were makes them
indistinguishable from the spectrum of corresponding Hermitian
Hamiltonian $H_0$ we found
 example, in which the energy of fermions is clearly dependent
on non-Hermitian characteristics. We are talking about the
consideration of the interaction of AMM of fermions with a
magnetic field. In this case we obtained the exact solution for
the energy of fermions (see (\ref{E61})).

It should be noted that the formula (\ref{E61})  is a valid not
only for charged fermions, but and for the neutral particles
possessing AMM. In this case one must simply replace the value of
quantized transverse momentum of a charged particle in a magnetic
field on the ordinary value $2\gamma n\rightarrow
{p_1}^2+{p_2}^2$.
 Thus, for the case of ultra cold polarized ordinary electronic
 neutrino with precision not over then linear
field terms   we can write

\be\label{E34} E(-1,0,0,H,m_{\nu_e}/M \rightarrow 0)= m_{\nu_e}
\sqrt{1-\frac{\mu_{\nu_e}}{\mu_0}\frac{
 H}{ H_c}}.
 \ee
However, in the case of exotic electronic
 neutrino we have

 \be\label{E3}
E(-1,0,0,H,m_{\nu_e}/M)= m_{\nu_e}
\sqrt{1-\frac{\mu_{\nu_e}}{\mu_0}\frac{2 M
 H}{m_{\nu_e} H_c}}.
 \ee

It is well known  \cite{n3},\cite{n4} that in the minimally
extended  SM  the one-loop radiative correction generates neutrino
magnetic moment which is proportional to the neutrino mass
\be\label{mu1}
  \mu_{\nu_e}=\frac{3}{8\sqrt{2}\pi^2}|e| G_F
  m_{\nu_e}=\left(3\cdot10^{-19}\right)\mu_0\left(\frac{m_{\nu_e}}{1
  eV}\right),
\ee where $ G_F$-Fermi coupling constant and $\mu_0$ is Bohr
magneton.
 However, so far, the most stringent laboratory constraints on the
 electron neutrino magnetic moment come from elastic
 neutrino-electron scattering experiments:
$ \mu_{\nu_e} <(1.5\cdot 10^{-10})\mu_0$\cite{n1}. Besides the
discussion of problem of measuring the mass of neutrinos (either
active or sterile) show that for an active neutrino model we have
$\sum m_\nu =0.320 eV$, whereas for a sterile neutrino $\sum m_\nu
=0.06 eV$ \cite{n2}.

One can also estimate the change of the border of region of
unbroken ${\cal P}{\cal T}$-symmetry due to the shift of the
lowest-energy state in the magnetic field. Using formulas
(\ref{E34}) and (\ref{E3}) we obtain correspondingly regions of
undisturbed  ${\cal P}{\cal T}$-symmetry  in the form

\be\label{border} H_{\nu_e -ordinary}\leq\frac{{\mu_0}}{
\mu_{\nu_e}} H_c ;\ee

\be\label{border1} H_{\nu_e -exotic}\leq\frac{m_{\nu_e}{\mu_0}}{2
M \mu_{\nu_e}} H_c. \ee

 Indeed let us take  the
following parameters of neutrino: the mass of the electronic
neutrino is  equal to $m_{\nu_e} = 1 eV$ and magnetic moment equal
to (\ref{mu1}). If we assume that the values of mass and magnetic
moment of exotic neutrino identical to parameters of ordinary
neutrinos, we can obtain the estimates of the border area
undisturbed ${\cal P}{\cal T}$ symmetry  for (\ref{border}) in the
form \be\label{E4}
 {H^{cr}}_{\nu_e -ordinary}  =  \frac{\mu_0}{\mu_{\nu_e}} H_c \sim 10^{32} Gauss.\ee
However in the case (\ref{border1}) the situation may change
radically \be\label{E41}
 {H^{cr}}_{\nu_e -exotic}  =  \frac{\mu_0}{\mu_{\nu_e}} \frac{m_{\nu_e}}{2 M} H_c \sim 10^4 Gauss.\ee

In comparison with (\ref{E4}) where the experimentally possible
field corrections are extremely small one can see that the
critical value of magnetic field (\ref{E41}) is attainable in the
sense of ordinary terrestrial experiments. In (\ref{E4}) and
(\ref{E41}) we used the values of quantum-electrodynamic constant
$H_c=4.41\cdot 10^{13}$Gauss and the Planck mass $M =
m_{Planck}\simeq 10^{19}GeV$. We do not know if there is an upper
limit to spectrum masses of elementary particles Ref.\cite{Mar}
and many are skeptical about the significance of the Planck mass,
but experimental studies of this thesis at high energies hardly
today may even be discussed. However contemporary precision of
alternative laboratory measurements at low energy in the magnetic
field may in principle allow to achieve the required values for
exotic particles in the near future. Thus, the obtained formulas
(\ref{E3})-(\ref{E41}) ensures possibility not only to be
convinced in the existence of the Maximal Mass but and in reality
of the so-called \emph{exotic particles}, because this phenomena
are inextricably related.

Note also that intensive magnetic fields exist near and within a
number of space objects. So, the magnetic fields intensity of the
order of $10^{12}\div10^{13}$Gauss observed near pulsars. Here
also may be  included the recent opening of such objects as
sources soft repeated gamma-ray burst and anomalous x-ray pulsars.
For them magneto-rotational models are proposed, and they were
named as magnetars. It was showed that for such objects achievable
magnetic fields with intensity up to $10^{15}$Gauss. It is very
important that the share of magnetars in the General population of
neutron stars reaches 10\%. In this regard, we note that the
processes with the participation of ordinary neutrinos and
especially of their possible "exotic partners in the presence of
such strong magnetic fields can have a significant influence on
the processes which may determine the evolution of astrophysical
objects.

 {\bf
Acknowledgment:} We are grateful to Prof. V.G.Kadyshevsky for
fruitful and highly useful discussions.

 \end{document}